\documentclass[aps,pra,twocolumn,groupedaddress,showpacs]{revtex4}

\usepackage{amsmath}
\usepackage{amsfonts}
\usepackage{amssymb}
\usepackage{graphics}
\usepackage{epsfig}
\usepackage{amsmath}

\begin{document}

\title{Quantum-limited force measurement with an optomechanical device}

\author{Marco Lucamarini}
\email{marco.lucamarini@unicam.it}
\author{David Vitali}
\author{Paolo Tombesi}
\affiliation{Dipartimento di Fisica, Universit\`a di Camerino,
I-62032 Camerino, Italy}

\date{\today}

\begin{abstract}
We study the detection of weak coherent forces by means of an
optomechanical device formed by a highly reflecting isolated mirror
shined by an intense and highly monochromatic laser field. Radiation
pressure excites a vibrational mode of the mirror, inducing
sidebands of the incident field, which are then measured by
heterodyne detection. We determine the sensitivity of such a scheme
and show that the use of an entangled input state of the two
sideband modes improves the detection, even in the presence of
damping and noise acting on the mechanical mode.
\end{abstract}


\maketitle

\section{Introduction}

Optomechanical systems play a crucial role in a variety of precision
measurement like gravitational wave detection \cite{grav} and atomic
force microscopes \cite{afm}. These systems are based on the
interaction between a movable mirror, the \textit{probe}
experiencing the force to be measured, and a radiation field, the
\textit{meter} reading out the mirror's position, which is due to
the radiation pressure force acting on the mirror. The mechanical
force exerts a momentum and position shift of a given vibrational
mode of the mirror, which in turn induces a phase shift of the
reflected optical field. A phase-sensitive measurement of the
reflected light provides therefore a measurement of the force.

These optomechanical force detectors have a sensitivity which is
limited by the thermal noise acting on the mirror mechanical degrees
freedom, as well as by the more fundamental, unavoidable, quantum
noise associated with the quantum nature of light, i.e., the phase
fluctuations of the incident laser beam (shot noise) and the
radiation pressure noise, inducing unwanted fluctuations of the
mirror position. A compromise between these noises leads to the
so-called standard quantum limit (SQL) for the sensitivity of the
measurement \cite{caves,BK92}. Analogous fundamental limitations
affect also other similar detection devices, such as nano- and
micro-electromechanical systems, which are also extensively studied
for the realization of ultra-sensitive detection devices
\cite{nems} such as force detection on the atto-Newton level
\cite{RUK} and mass detection on the zepto-gram level \cite{SER}.

Many proposals for the detection of weak forces involves
high-finesse optical cavities with a movable mirror, in which the
phase sensitivity is proportional to the cavity finesse
\cite{qlock}. However, recently Ref.~\cite{FMT04} has proposed a new
optomechanical detection scheme involving a \emph{single} highly
reflecting mirror, shined by an intense highly monochromatic laser
pulse. A vibrational mode of the mirror induces two sidebands of the
incident field, the Stokes and anti-Stokes sideband. This effect was
recently observed in a micro-mechanical resonator as a consequence
of the radiation pressure force acting on it \cite{VAHA}. Under
appropriate conditions on the duration of the laser pulse, the two
sideband modes show significant two-mode squeezing, i.e., they are
strongly entangled \cite{PM+03}. In particular the difference
between the two amplitude quadratures and the sum of the phase
quadratures of the sideband modes can be highly squeezed
\cite{pirjmo}, and this reduced noise properties are used in
Ref.~\cite{FMT04} to achieve high-sensitive detection of a force
acting on the mirror. However Ref.~\cite{FMT04} considered only
partially the effect of the thermal environment of the mechanical
mode. In fact, Ref.~\cite{FMT04} considered the limiting case of a
laser pulse duration much shorter than the mechanical relaxation
time and neglected all the dynamical effects of damping and thermal
noise. Here we drop this assumption and we take into account the
effects of the thermal environment acting on the mechanical mode, by
adopting a quantum Langevin equation treatment \cite{Gard91}. We
shall see that, as expected, damping and thermal noise have a
detrimental effect on the force detection sensitivity, but that one
can still go below the SQL at achievable values of mechanical
damping and temperatures, provided that the two sideband modes are
appropriately entangled at the input.

The outline of the paper is as follows. In Sec. II we illustrate the
model describing the force detection scheme, while in Sec. III we
define and evaluate the minimum detectable force. In Sec. IV we
consider experimentally achievable parameters and compare the
performance of the scheme with the SQL for the detection of a force
\cite{BK92}, while Sec. V is for concluding remarks.

\section{The Model}

\begin{figure}[ht]
  \includegraphics[width=.50\textwidth]{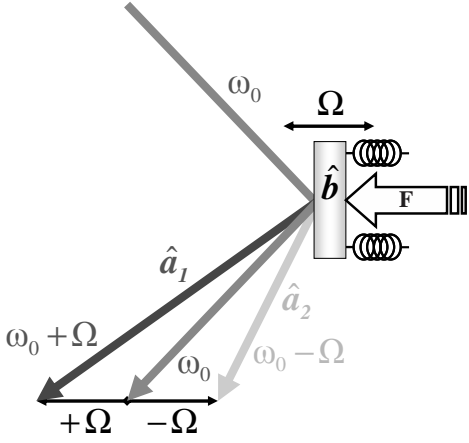}\\
  \caption{Schematics of the opto-mechanical device to detect a
  force F. A vibrational
  mode of the mirror (\textit{probe}) related to the ladder
  operator $\hat{b}$ and oscillating at angular frequency  $\Omega$
  is excited by the radiation pressure of the incident laser field
  (angular frequency $\omega_0$). Light is scattered into the two first sideband
  modes, the
  anti-Stokes mode at $\omega_0+\Omega$ (operator $\hat{a}_{1}$) and
  the Stokes mode at $\omega_0-\Omega$ (operator $\hat{a}_{2}$).
  The force F to be
  detected acts on the mirror along the horizontal direction. } \label{Setup}
\end{figure}

We consider the system schematically depicted in Fig.~\ref{Setup}.
It consists of a perfectly reflecting mirror shined by a pulsed
quasi-monochromatic laser at main frequency $\omega_0$, linearly
polarized in the mirror surface and focused in such a way as to
excite the Gaussian acoustic modes of the mirror, in which only a
small portion of the mirror around its center vibrates.
These modes describe elastic deformations of the mirror along the
direction orthogonal to the surface, and are characterized by a
small waist $w$, a large mechanical quality factor $Q$ and a small
effective mass $M$. The motion of the mirror is actually determined
by the excitation of several modes with different resonant
frequencies. However, a single frequency mode can be considered when
a bandpass filter in the detection scheme is used \cite{Pinard} and
mode-mode coupling is negligible. Therefore we will consider a
single mechanical mode of the mirror only, which can be modeled as
an harmonic oscillator with mass equal to the effective mass $M$ and
angular frequency $\Omega$,
\begin{equation}
H_{0}=\frac{P_{b}^{2}}{2M}+\frac{1}{2}M\Omega^{2}
X_{b}^{2},\label{H_0}
\end{equation}
where $X_{b}$ and $P_{b}$ are position and momentum operators of the
chosen Gaussian vibrational mode, satisfying the commutation rule $
[ X_{b},P_{b}]=i\hbar $.

As demonstrated in Ref.~\cite{PM+03}, in the case of an intense
classical incident laser field and neglecting fast terms oscillating
at $\pm \Omega$, the interaction between the chosen vibrational mode
and the continuum of electromagnetic modes can be written, in the
interaction picture (IP) with respect to the free Hamiltonian of the
system, as a simple bilinear Hamiltonian involving the vibrational
mode and the two first optical sideband modes
\begin{equation}
\tilde{H}_{int} =i\hbar\chi\left(\tilde{a} _{1}^{+}\tilde{b}^{+}-
\tilde{a}_{1}\tilde{b}\right)
+i\hbar\theta\left(\tilde{a}_{2}^{+}\tilde{b}- \tilde{a}
_{2}\tilde{b}^{+}\right) ,\label{H_int_2}
\end{equation}
where $b$ is the annihilation operator of the vibrational mode given
by $b=\left(iP_b+M\Omega X_b\right)/\sqrt{2\hbar M\Omega}$, $a_{k}$
are the annihilation operators of the optical sidebands Stokes
($k=1$, angular frequency $\omega_0-\Omega$) and anti-Stokes ($k=2$,
angular frequency $\omega_0+\Omega$) modes, and the tilded operators
are those in the IP. The coupling constants in Eq.~(\ref{H_int_2})
are given by \cite{PM+03}
\begin{eqnarray}
\label{chi} \chi &=&
\cos\phi_{0}\sqrt{\frac{{\wp}\Delta\nu_{det}^{2}(\omega_{0}-\Omega)}{2M\Omega
c^{2}\Delta\nu_{mode}}}, \\
\theta &=& \chi \sqrt{\frac{\omega_{0}+\Omega}{\omega_{0}-\Omega}},
\end{eqnarray}
where $\phi_{0}$ is the angle of incidence of the driving beam,
$\wp$ is the power of the incident beam and $ \Delta\nu_{mode}$ is
its bandwidth, while $ \Delta\nu_{det}$ is the detection bandwidth.

\subsection{Including damping and thermal noise on the vibrational motion}

The performance of the three-mode optomechanical system described by
Eq.~(\ref{H_int_2}) as a detector of a weak classical force acting
on the mirror has been studied in Ref.~\cite{FMT04}. In this latter
paper, dynamical effects of the thermal environment of the mirror
were ignored, because the dynamics have been studied only in the
limit of interaction times much shorter than the mechanical damping
time. Here we consider the more realistic situation of
non-negligible mechanical damping, which, due to the
fluctuation-dissipation theorem, implies also considering the effect
of thermal noise on the mirror vibrational mode. These effects are
described in terms of a quantum Langevin equation (QLE) for the
vibrational mode which, in the first Markov approximation, can be
written as~\cite{Gard91}
\begin{equation}
\overset{\cdot}{Y}=\frac{i}{\hbar}\left[H_0+  H_{int},Y\right]
-\frac{i}{2\hbar }\left\{  \left[  X_{b},Y\right]
,\xi(t)-\eta\overset{\cdot}{X} _{b}\right\} , \label{Langevin}
\end{equation}
where the brackets $\{,\}$ represent the anti-commutator, $Y$ is a
generic Heisenberg-picture operator, $\eta$ is the damping
coefficient of the mirror and $\xi(t)$ is the stochastic noise
force, with correlation function \cite{Gard91,GV01}
\begin{equation}\label{browncorre}
\left \langle \xi(t) \xi(t')\right \rangle = \hbar \eta
\int_{-\infty}^{\infty} \frac{d\omega}{2\pi} e^{-i\omega(t-t')}
\omega \left[\coth\left(\frac{\hbar \omega}{2k_BT}\right)+1\right],
\end{equation}
where $k_B$ is the Boltzmann constant and $T$ is the equilibrium
temperature. Applying Eq.~(\ref{Langevin}) for the mirror position
and momentum operators, we get
\begin{align*}
\overset{\cdot}{X_{b}} &  =\frac{P_{b}}{M}+i[H_{int},X_{b}]\\
\overset{\cdot}{P_{b}} & =-M\Omega^{2}X_{b}+i[H_{int},P_{b}]
-\overset{\cdot}{\eta X_{b}}+\xi(t),
\end{align*}
implying the following evolution equation for $b$:
\begin{eqnarray}
\overset{\cdot}{b} & =& -i\Omega
b+\frac{i}{\hbar}[H_{int},b]-\frac{\eta}{2M}\left(
b-b^{+}\right) \nonumber \\
&&  +i\sqrt{\frac{1}{2M\hbar\Omega}}\xi(t)+\frac{\eta}{2\hbar
M\Omega }[H_{int},b+b^{+}]. \label{b}
\end{eqnarray}
If we now move to the IP we get
\begin{align}
\overset{\cdot}{\tilde{b}}  & =\frac{i}{\hbar}[\tilde{H}
_{int},\tilde{b}]-\frac{\eta}{2M}\left( \tilde{b}-\tilde{b}
^{+}\right) \nonumber \\
& +i\sqrt{\frac{1}{2M\hbar\Omega}}e^{i\Omega t}\xi(t)+\frac{\eta}
{2M\hbar\Omega}[\tilde{H}_{int},\tilde{b}+\tilde{b}^{+}].
\label{b_tilde}
\end{align}
By virtue of Eq.~(\ref{H_int_2}) we have:
\[
\frac{\tilde{H}_{int}}{\hbar\Omega}\simeq\frac{\chi}{\Omega}\simeq
\frac{\theta}{\Omega}.
\]
This ratio is usually very low when realistic values are taken
into account ($\chi\simeq\theta\sim10^{5}$ s$^{-1}$,
$\Omega\sim10^{8}$ s$^{-1}$) \cite{PM+03}. Then the last term of
Eq.~(\ref{b_tilde}) is much smaller than the second term, and can
be neglected. Moreover, the term $\tilde{b}^{+}$ is
counter-rotating and since we have already neglected fast
oscillating terms at the frequency $\Omega$, we have to neglect it
in Eq.~(\ref{b_tilde}) for consistency. In this way we arrive at
the final quantum Langevin equation for the vibrational mode in
the IP
\begin{equation}
\overset{\cdot}{\tilde{b}}=\frac{i}{\hbar}\left[ \tilde{H}
_{int},\tilde{b}\right] -2\gamma
\tilde{b}+2\sqrt{\gamma}\tilde{b}_{in}, \label{Langevin_b}
\end{equation}
where we have defined the damping rate $\gamma$ and the scaled
noise $\tilde{b}_{in}(t)$ as
\begin{eqnarray}
\gamma & =& \frac{\eta}{4M},\\
\tilde{b}_{in}(t) &=& i\frac{e^{i\Omega t}\xi(t)}{\sqrt{2\eta
\hbar\Omega} }.
\end{eqnarray}
In the limit of large $\Omega$ we are considering, the correlation
functions of this latter noise term becomes simple. In fact, using
Eq.~(\ref{browncorre}) and the fact that the factor $e^{i\Omega t}$
is rapidly oscillating within the timescales of interest, it is
possible to derive the following correlation properties of
$\tilde{b}_{in}(t)$ \cite{Gard91-bis},
\begin{subequations}
\label{correcm}
\begin{eqnarray}
\langle
\tilde{b}_{in}(t)\tilde{b}_{in}(t')\rangle&=&0,\\
\langle
\tilde{b}_{in}(t)\tilde{b}_{in}^{\dagger}(t')\rangle&=&(1+\overline{n})\delta(t-t'),\\
\langle \tilde{b}_{in}^{\dagger}(t)\tilde{b}_{in}(t')\rangle &=&
\overline{n}\delta(t-t'),
\end{eqnarray}
\end{subequations}
where $\overline{n}=1/(e^{\hbar\Omega/k_BT}-1)$ is the mean thermal
vibrational number at the equilibrium temperature $T$. To state it
in an equivalent way, in the limit of large $\Omega$, the properties
of the Brownian noise acting on the vibrational mode become similar
to those of the input noise of optical systems.

\subsection{Exact solution of the dynamics}
\label{subsec:ex_sol}

The Stokes and anti-Stokes modes are not directly sensitive to the
damping and noise acting on the mirror. Moreover they do not undergo
additional loss mechanisms since they are traveling waves.
Therefore, the Heisenberg-Langevin equations describing the dynamics
of the whole system, in the presence of an additional constant force
acting on the mirror with dimensionless strength $f$, are
\begin{subequations}
\label{hei-lan}
\begin{eqnarray}
\dot{\tilde{a}}_{1} &  = &\chi \tilde{b}^{\dagger} \\
\dot{\tilde{b}} &  = & \chi \tilde{a}_{1}^{\dagger}-\theta
\tilde{a}_{2}-2\gamma
\tilde{b}+2\sqrt{\gamma}\tilde{b}_{in}+i\Omega fe^{i\Omega t}\\
\dot{\tilde{a}}_{2} &  = &\theta \tilde{b}.
\end{eqnarray}
\end{subequations}
From these we get the equation for $\tilde{b}$ alone
\begin{equation}
\ddot{\tilde{b}}(t)+2\gamma\dot{\tilde{b}}(t)+\Theta^{2}\tilde{b}(t)=g(t),
\end{equation}
where
\begin{eqnarray}
&& \Theta= \sqrt{\theta^2-\chi^2} \\
&& g(t)=-\Omega^{2}fe^{i\Omega
t}+2\sqrt{\gamma}\overset{\cdot}{\tilde{b}}_{in}(t).
\end{eqnarray}
After a straightforward calculation the solution for $\tilde{b}$
reads:
\begin{align}
\tilde{b}(t) &
=\frac{\chi}{\omega}S(t)a_{1}^{\dagger}(0)-\frac{\theta}{\omega}
S(t)a_{2}(0)+\left(  C(t)-\frac{\gamma}{\omega}S(t)\right)  b(0) \nonumber \\
&  +\Omega F_{+}\left[  C(t)-\left(
\frac{\gamma}{\omega}-i\frac{\Theta^{2}
}{\omega\Omega}\right)  S(t)-e^{i\Omega t}\right] \nonumber \\
&  +\int_{0}^{t}dsK\left( t-s\right)  \tilde{b}_{in}(s)
\label{b(t)}
\end{align}
where
\begin{align*}
\omega & =\sqrt{\Theta^2-\gamma^2} \\
S(t)  & =e^{-\gamma t}\sin\omega t\\
C(t)  & =e^{-\gamma t}\cos\omega t\\
F_{\pm}  & =\frac{\Omega f}{\Theta^{2}-\Omega^{2}\pm2i\gamma\Omega}\\
K\left(t-s\right)    & =\frac{\sqrt{\gamma}}{\omega}\left[ i\left(
\gamma-i\omega\right)  e^{\left(i\omega-  \gamma\right) \left(
t-s\right) }+c.c.\right]
\end{align*}
We have assumed $\omega$ real, i.e., $\gamma < \Theta$, which is
typically satisfied in the experimentally relevant limit of a high-Q
vibrational mode. Notice that we reobtain the results of
Ref.~\cite{FMT04} in the limit $\gamma=0$. The exact expressions for
the optical sidebands annihilation operators are instead given by
\begin{align}
a_{1}(t)  & =\frac{1}{\Theta{}^{2}}\left[
\theta^{2}-\chi^{2}\allowbreak \left(
C(t)+\frac{\gamma}{\omega}S(t)\right)  \right] a_{1}(0)+\frac{\chi
}{\omega} S(t) b(0)^{\dagger} \nonumber \\
& +\frac{\chi\theta}{\Theta^{2}}\left[  \allowbreak
C(t)+\frac{\gamma
}{\omega}S(t)-1\right]  a_{2}^{\dagger}(0) \nonumber \\
& +i\chi F_{-}\left[  C(t)+\left(
\frac{\gamma}{\omega}-i\frac{\Omega
}{\omega}\right)  S(t)-e^{-i\Omega t}\right] \nonumber \\
& +\chi\int_{0}^{t}dt^{\prime}\int_{0}^{t^{\prime}}ds K\left(
t^{\prime}-s\right) \tilde{b}_{in}^{\dagger}(s), \label{a1(t)}
\end{align}
\begin{align}
a_{2}(t)  & =\frac{\chi\theta}{\Theta{}^{2}}\left[  1-\allowbreak
C(t) -\frac{\gamma}{\omega}S(t)\right]
a_{1}^{\dagger}(0)+\frac{\theta}{\omega
}S(t)  b(0) \nonumber \\
& +\frac{1}{\Theta{}^{2}}\left[  \theta^{2}\left(
C(t)+\frac{\gamma}{\omega
}S(t)\right)  -\chi^{2}\right]  a_{2}(0) \nonumber \\
& -i\theta F_{+}\left[  C(t)+\left(
\frac{\gamma}{\omega}+i\frac{\Omega
}{\omega}\right)  S(t)-e^{i\Omega t}\right] \nonumber \\
& +\theta\int_{0}^{t}dt^{\prime}\int_{0}^{t^{\prime}}dsK\left(
t^{\prime}-s\right) \tilde{b}_{in}(s). \label{a2(t)}
\end{align}

\section{Force detection sensitivity}

We now consider the real-time detection of the constant force $f$
applied to the mirror and determine the sensitivity of the
considered optomechanical system, by evaluating the corresponding
signal-to-noise ratio. In optomechanical devices based on radiation
pressure effects, one typically performs phase-sensitive
measurements on the reflected beam (the meter) because the force to
be detected shifts the mechanical probe determining in this way a
phase-shift of the field.

As suggested in \cite{FMT04}, we consider an appropriate heterodyne
detection~\cite{YS80} of the two sideband modes, corresponding to
the measurement of the operator
\begin{align*}
Z_{\varphi}  & =e^{i\varphi}a_{1}-e^{-i\varphi}
a_{2}^{\dagger}\\
&  =\cos\varphi\left( a_{1}-a_{2}^{\dagger}\right)
+i\sin\varphi\left( a_{1}+a_{2}^{\dagger}\right) ,
\end{align*}
where $\varphi$ is an experimentally adjustable phase. We choose
$\varphi=\pi$ and consider in particular the imaginary part of
$Z_{\pi}$,
\begin{equation}
Z_{\pi}^{I}
=\frac{Z_{\pi}-Z_{\pi}^{\dagger}}{2i}=\frac{a_{1}^{\dagger}-a_{1}+a_{2}^{\dagger}-a_{2}}{2i}.
\label{Z_pi_i}
\end{equation}
Using Eqs.~(\ref{b(t)})-(\ref{Z_pi_i}) one gets
\begin{eqnarray}
&& \frac{Z_{\pi}^{I}(t)}{\chi-\theta} =A_{1}\left(  t\right)
Y_{1}(0)+A_{2}\left(  t\right)
Y_{2}(0) +B\left(  t\right)  Y_{b}(0)\nonumber \\
&& +G(t) +\int_{0}^{t}dt^{\prime}\int_{0}^{t^{\prime}}ds
D\left(t^{\prime}- s\right)  Y_{b_{in}}(s),  \label{Z_obs}
\end{eqnarray}
where $Y_{k}(0)=[a_{k}(0)-a_{k}^{\dagger}(0)]/2i$, $k=\{1,2,b\}$,
$Y_{b_{in}}(s) =[b_{in}(s)-b _{in}^{\dagger}(s)]/2i$,
\begin{subequations}
\label{defin}
\begin{eqnarray}
A_{1}\left(  t\right)   &  = &\frac{\theta+\chi
\Upsilon_{+}(t)}{\Theta^{2}}, \\
B\left(  t\right)   &  = &\frac{S(t)}{\omega},\\
A_{2}\left(  t\right)   &  =& \frac{\chi+\theta
\Upsilon_{+}(t)}{\Theta^{2}}, \\
G\left(  t\right)   &  = & \frac{\left\vert F_{\pm}\right\vert
^{2}}{\Omega f}\left\{\left[\Upsilon_{+}(t)-\cos\Omega t\right]
\left(
\Omega^{2}-\Theta^{2}\right) \right. \nonumber \\
&-& \left. 2\gamma\Omega\left[
\frac{\Omega}{\omega}S(t)-\sin\Omega t\right]\right\},\\
D(t)   &  = & 2\sqrt{\gamma} \Upsilon_{-}\\
\Upsilon_{\pm}(t)  &  = & C(t)\pm\frac{\gamma}{\omega}S(t).
\end{eqnarray}
\end{subequations}
The signal is given by the absolute value of the mean value of the
observed quantity, i.e., $\mathcal{S}=\left|\left\langle Z_{\pi}
^{I}(t)\right\rangle \right|$, while the noise corresponds to the
square root of the variance of the same observable, $\mathcal{N}
=\left\langle Z_{\pi}^{I}(t)^{2}\right\rangle -\left\langle
Z_{\pi}^{I}(t)\right\rangle ^{2}$. Since we are considering an open
system, averaging means taking expectation values with respect to
the initial state of the system \emph{and} the environment. In the
QLE treatment this means averaging over the initial state of our
optomechanical system and over the noise $b_{in}(t)$.

The natural initial state of the optomechanical system is the
product state $\rho_{tot}(0)=|0\rangle_1 \langle 0 | \otimes
|0\rangle_2 \langle 0 |\otimes \rho_{th}^b$, where the two sideband
modes are in the vacuum state and the vibrational mode is in thermal
equilibrium with mean vibrational number $\overline{n}$,
\begin{equation} \label{initherm}
\rho_{th}^{b}=\sum_{n}\frac{\overline{n}^{n}}{\left(
1+\overline{n}\right) ^{n+1}}\left\vert n\right\rangle \left\langle
n\right\vert .
\end{equation}
However, as suggested in \cite{FMT04,holl}, the use of nonclassical
states, and in particular \emph{entangled} states of the optical
modes, could improve force detection sensitivity. For this reason we
consider the following class of pure initial states for the two
sideband modes,
\begin{equation}
\left\vert \Psi\right\rangle
_{12}=\sqrt{1-\tanh^{2}s}\sum_{n=0}^{\infty }\left(  \tanh s\right)
^{n}\left\vert n\right\rangle _{1}\left\vert n\right\rangle _{2},
\label{initial_state}
\end{equation}
with $s\in\mathbb{R}$, that is, a two-mode squeezed state,
reproducing the usual vacuum state initial condition for $s=0$ and
showing entanglement between the two optical sidebands whenever $s
\neq 0$. Notice that when $s \neq 0$, a nonzero incident light power
is present not only at the carrier frequency $\omega_0$ ($\wp_0$),
but also at the two sideband frequencies $\omega_0 \pm \Omega$
($\wp_{1,2}$), because power is proportional to $\sinh ^2 s$
\cite{MW94}. Therefore, if $s$ is sufficiently large, one could have
non-negligible scattered light at the additional sideband
frequencies $\omega_0 \pm 2\Omega$ and interference effects at
$\omega_0$. This however happens only at unrealistically large
values of two-mode squeezing $s$. Therefore, we consider not too
large values of $s$, so that $\wp_{1,2} \ll \wp_0 \sim \wp$ and
neglect these additional effects.

Using the initial conditions (\ref{initherm}) and
(\ref{initial_state}), and the fact that $\langle b_{in}(t)\rangle
=0$, one gets that the signal can be written as:
\begin{equation}
\mathcal{S}= \left\vert \left(\theta-\chi\right)
G(t)\right\vert,
 \label{Signal_S}
\end{equation}
where $G(t)$ is given by Eq.~(\ref{defin}d), while the noise is
given by the square root of the following variance:
\begin{eqnarray}
&& \mathcal{N}  =\left(\theta-\chi\right)
^{2}\left\{A_{1}^{2}(t)\langle Y_{1}(0)^{2}\rangle +B^{2}(t) \langle
Y_{b}(0)^{2}\rangle \right. \nonumber\\
&& \left. +A_{2}^{2}(t)\langle Y_{2}(0)^{2}\rangle +2A_{1}(t)
A_{2}(t) \langle Y_{1}(0)Y_{2}(0)\rangle \right. \nonumber \\
&& \left. +\left\langle \left(
\int_{0}^{t}dt^{\prime}\int_{0}^{t^{\prime}} ds D( t^{\prime}-s)
Y_{b_{in}}(s)  \right) ^{2}\right\rangle \right\}. \label{Noise_N}
\end{eqnarray}
If we compare these results with the corresponding ones of
Ref.~\cite{FMT04}, which considered the same force detection scheme,
but in the limit $\gamma \to 0$ (which means neglecting the
dynamical effects of the environment), we see that the signal and
noise have the same structure, with two important differences. First
of all, the time-dependent coefficients $A_1(t)$, $A_2(t)$, $B(t)$,
$G(t)$ have a modified expression due to the nonzero damping rate
$\gamma$; moreover in the present case, the noise has an additional
term, corresponding to the last line of Eq.~(\ref{Noise_N}). Using
the definition of $Y_{b_{in}}(s)$ and the correlation functions of
Eqs.~(\ref{correcm}), one gets $ \langle Y_{b_{in}}(s)
Y_{b_{in}}(s^{\prime })\rangle =(2\overline{n}+1)\delta
(s-s^{\prime})/4 $, so that the explicit expression of this
additional noise term is:
\begin{eqnarray}
&& \left\langle \left(
\int_{0}^{t}dt^{\prime}\int_{0}^{t^{\prime}} ds D( t^{\prime}-s)
Y_{b_{in}}(s)  \right) ^{2}\right\rangle = \nonumber \\
&&=\frac{\left( 2\overline{n}+1\right)}{4}\left[
\frac{1}{\Theta^{2}}-\frac{1}{\omega^{2}}e^{-2\gamma
t}+\frac{\gamma^{2}}{\omega^{2}\Theta^{2}}
C(2t)-\frac{\gamma}{\omega\Theta^{2}} S(2t)\right] \nonumber \\
&&= \frac{\left( 2\overline{n}+1\right)}{4}~E^{2}(t),
\label{ThermNoise}
\end{eqnarray}
which is a positive, non-decreasing function of time for any
positive $t$. By using Eqs.~(\ref{Signal_S})-(\ref{ThermNoise})
and the following initial mean values, stemming from
Eqs.~(\ref{initherm})-(\ref{initial_state}),
\begin{subequations}
\label{inimean}
\begin{eqnarray}
\langle Y_{1}(0)^{2} \rangle & =&
\langle Y_{2}(0)^{2} \rangle =\frac{ 1+2\sinh^{2}s}{4},  \\
\langle Y_{b}(0)^{2} \rangle  & = & \frac{ 2\overline{n}+1}{4}, \\
\langle Y_{1}(0)Y_{2}(0) \rangle  & = & -\frac{\sinh2s}{4},\\
\langle Y_{1}(0)Y_{b}(0) \rangle   & = & \langle
Y_{2}(0)Y_{b}(0)\rangle  =0,
\end{eqnarray}
\end{subequations}
one can obtain the explicit expression of the signal-to-noise ratio
$\mathcal{R} = \mathcal{S}/\sqrt{\mathcal{N}}$. In order to have
significant results $\mathcal{R}$ must be greater than a certain
confidence level $\alpha$. This parameter is fixed by the
experimenter in accordance to his trust in the measuring device; for
simplicity we set here $\alpha=1$. The \textit{sensitivity} or
\textit{minimum detectable input} of the device is the minimum
magnitude of the input signal required to produce an output with a
specified signal-to-noise ratio. It is easy to see that in order to
obtain $\mathcal{R} =1$ the sensitivity $f_{min}$ of the apparatus
of Fig.~\ref{Setup} must be at least equal to
$\sqrt{\mathcal{N}}/\mathcal{S}$. This provides the following
explicit expression of the minimum force detectable with the
apparatus at issue:
\begin{eqnarray}
&& f_{\min}  =\frac{\left(\Theta^{2}-\Omega^{2}\right) ^{2}+4\gamma
^{2}\Omega^{2}}{2\Omega\left\vert \Delta(t)\right\vert } \nonumber \\
&& \times \left\{  \left[ A_{1}(t) \cosh s  -A_{2}(t)\sinh s\right]^{2} \right. \nonumber \\
& & \left. +\left[  A_{1}(t)\sinh s  -A_{2}(t)\cosh s\right]^{2} \right.\nonumber \\
& & \left. +\left(  2\overline{n}+1\right)  \left[  B^{2}\left(
t\right) +E^{2}(t)\right]  \right\}^{\frac{1}{2}}, \label{fmin_2}
\end{eqnarray}
where
\begin{eqnarray}
\Delta(t)  &=& \left(\Omega^{2}-\Theta^{2}\right)  \left[
C(t)+\frac{\gamma }{\omega}S(t)-\cos\Omega t\right] \nonumber
\\ & - & 2\gamma\Omega\left [\frac{\Omega}{\omega}S(t)-\sin\Omega
t\right].
\end{eqnarray}

Let us note that $f$ and $f_{\min}$ are dimensionless quantities.
The scaling factor to pass to a minimum detectable force with proper
dimensions $F$ can be obtained from Eq.~(\ref{hei-lan}b) and the
usual definition of the operator $\tilde{b}$, giving
\begin{equation}\label{factor}
    \frac{F}{f_{min}} = \Omega \sqrt{2\hbar M \Omega}.
\end{equation}

\section{Minimum detectable force and standard quantum limit}

In this section we study the performance of the force detection
scheme presented here, characterized by the minimum detectable
force, Eq.~(\ref{fmin_2}). In this respect a significative
benchmark is provided by the so called \textit{standard quantum
limit} (SQL), defined in~\cite{BK92}.

The SQL represents the optimal sensitivity pertaining to an ideal
quantum harmonic oscillator when used to measure a force applied
to it. One typically considers only the limitations due to quantum
uncertainties, i.e., those associated with the oscillator's
quantum ground state, and assumes zero temperature and no damping
(shot noise limit). Given an effective interaction time $\tau$
between the force and an oscillator of mass $M$ and angular
frequency $\Omega$, the SQL for the detection of a constant force
is given by~\cite{BK92}
\begin{equation}\label{SQL}
    F_{SQL} = \frac{\sqrt{\hbar \Omega M}}{\tau}.
\end{equation}
This means that in principle $F_{SQL}$ can even become zero when
$\tau$ tends to infinite. In practice however the interaction time
$\tau$ cannot be too large. In a realistic setup the mechanical
damping rate $\gamma$ is always nonzero and this fixes a first
upper limit, $\tau \ll 1/\gamma $. Moreover, a perfectly constant
force is just an idealization; usually one has some typical time
scale $\tau_F$ over which the force appreciably changes. As a
consequence, it is not convenient to take $\tau \gg \tau_F$
because in such a case the momentum change induced by the force
may average to zero; this fixes a further upper bound for $\tau$.
So while in Sec.~\ref{subsec:ex_sol} we have assumed a constant
force, this practically means that $f$ does not appreciably vary
over the typical timescale of the system dynamics, which is
essentially determined by $\Theta^{-1}$. Hence a reasonable choice
for $\tau$ is $\tau \sim \Theta^{-1}$, and we shall set
\begin{equation}\label{tau}
    \tau=2\pi/\Theta
\end{equation}
in the expression for the SQL, Eq.~(\ref{SQL}). Since
$\gamma\ll\Theta$, the above choice is consistent with the
requirement $\tau \ll 1/\gamma $. Furthermore, we show below that
this choice results optimal with respect to the final heterodyne
measurement of the sideband modes.

In order to compare the sensitivity of Eq.~\eqref{fmin_2}, scaled
with the factor of Eq.~\eqref{factor}, with the SQL for the
detection of a force, Eq.~\eqref{SQL}, we choose the parameter
regime illustrated in Table~\ref{tab1} which, even though
challenging to achieve, is within reach of current technology.
This parameter choice gives $ F_{SQL} =
12.2\times10^{-18}~\textrm{N} $.

\begin{table}[h!]
\centering
\begin{tabular}
[c]{|c|c|}\hline Parameter & Value\\\hline $2\pi c/\omega_{0}$ &
$600$ nm\\\hline $\Omega$ & $2\pi\times10^{7}$ Hz\\\hline $P$ &
$50$ mW\\\hline $M$ & $5\times10^{-12}$ Kg\\\hline
$\Delta\nu_{det}$ & $10^{6}$ Hz\\\hline $\Delta\nu_{mode}$ &
$10^{2}$ Hz\\\hline $\gamma$ & $1$ Hz\\\hline
\end{tabular}\label{tab1}
\caption{Choice of parameters for the detection scheme of
Fig.~\ref{Setup}.}\label{tab1}
\end{table}

\begin{figure}[ptb]
\begin{center}
\includegraphics[width=0.45\textwidth]{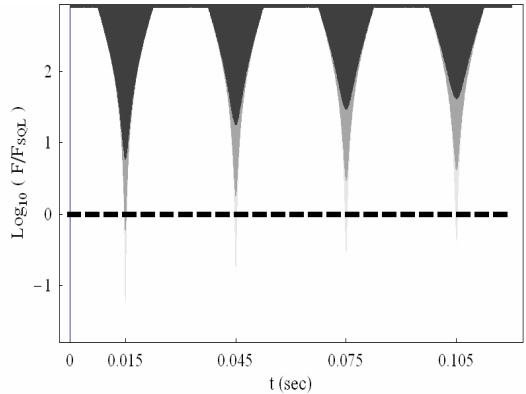}
\caption{Plot of the envelope of
$\textrm{Log}_{10}(\textrm{F}/\textrm{F}_{\textrm{SQL}})$ (rapidly
oscillating at the angular frequency $\Omega$) versus the
interaction time, at four different values of damping,
$\gamma=\{0.01,0.1,1\}$~Hz, corresponding to increasingly darker
grey curves. The other parameters are given by Table~1, while the
mirror temperature is T = 0 and there is no initial entanglement
between the sideband modes, $s$ = 0. The dashed line represents
the SQL.} \label{Fig_fmin_t}
\end{center}
\end{figure}

In Fig.~\ref{Fig_fmin_t} we plot
$\textrm{Log}_{10}(\textrm{F}/\textrm{F}_{\textrm{SQL}})$ as a
function of the interaction time $t$ (i.e., the duration of the
driving laser pulse), for different values of the damping rate,
$\gamma=\{0.01,0.1,1\}~\textrm{Hz}$, corresponding to increasingly
darker grey curves. The temperature and the two-mode squeezing
parameter $s$ are taken to be zero. We see that the minimum
detectable force is a quasi-periodic function with period
$2\pi/\omega \simeq 2\pi/\Theta$, presenting a series of minima at
times $t=\left( 2k+1\right) \pi/\omega $, with $k$ a positive
integer. This is due to the entanglement dynamically produced by
the interaction Hamiltonian of Eq.~\eqref{H_int_2}. In fact, the
minima are obtained when the two reflected sideband modes are
factorized from the vibrational mode and are in a two-mode
squeezed state in which the variance of the difference of the
amplitude quadratures of the two sideband modes, as well as the
variance of the sum of their phase quadratures, are maximally
squeezed below the shot noise limit ~\cite{FMT04,pirjmo}. Since
the measured observable, $Z_{\pi}^{I}$ of Eq.~(\ref{Z_pi_i}), is
just the sum of the phase quadratures of the two fields, the
minima corresponds to the minimum noise, yielding the maximum
sensitivity of the detection scheme.

As the interaction time increases, the minimum detectable force at
the local minima increases as well, because the effect of damping
becomes more and more important at longer times. As a consequence,
the first minimum at $t_1 \simeq \pi/\Theta$, ($t_1 \simeq 15$~ms
with the parameter values of Table~\ref{tab1}) corresponds to the
best possible sensitivity attainable with the device at issue. In
typical situations, the time of arrival of the slowly varying force
to be detected is unknown (consider for example the case of the
tidal force of a gravitational wave). In such a case, the best
detection strategy corresponds to a \emph{pulsed} regime, in which a
precise temporal switch fixes a pulse duration exactly equal to
$t_1$ for the impinging laser, with a repetition rate of the order
of the damping frequency $\gamma$.

The dashed line in Fig.~\ref{Fig_fmin_t} represents the SQL,
obtained when $\textrm{F}/\textrm{F}_{\textrm{SQL}}=1$. It is
apparent that for sufficiently low values of $\gamma$ the
sensitivity goes beyond the SQL. However, when $\gamma$ assumes the
more realistic value of $1$ Hz (darkest grey in the figure) the
entanglement created by the dynamics is no more sufficient to go
below the SQL, in none of its minima. However, as shown by
Ref.~\cite{FMT04}, there is a further resource that can be exploited
in order to beat the SQL, i.e., the two-mode squeezing of the
initial state of Eq.~\eqref{initial_state}. This factor represents a
sort of ``static'' entanglement that can add its effect to that of
the dynamically generated entanglement between the sidebands and is
able to increase significantly the sensitivity of the apparatus.

Hereafter we concentrate on the first minimum of the minimum
detectable force of Fig.~\ref{Fig_fmin_t}, that is we fix $t_1
\simeq \pi/\Theta$.
\begin{figure}[h!]
\vspace{0.5cm}
\begin{center}
\includegraphics[width=0.45\textwidth]{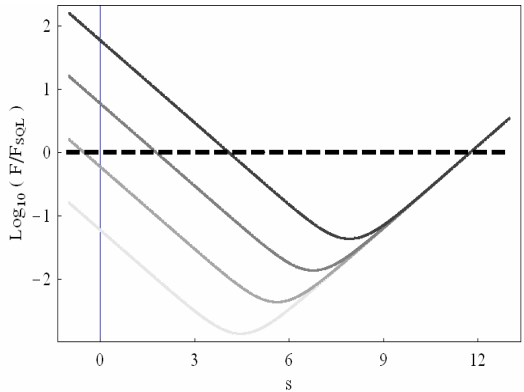}
\caption{Plot of
$\textrm{Log}_{10}(\textrm{F}/\textrm{F}_{\textrm{SQL}})$ versus
the two-mode squeezing factor $s$, at four different values of
damping, $\gamma = \{0.01, 0.1, 1, 10\}$~Hz, corresponding to
increasingly darker grey curves. The dashed line represents the
SQL. The other parameters are given by Table~1, while the mirror
temperature is T = 0.} \label{Fig_fmin_gam}
\end{center}
\end{figure}
In Fig.~\ref{Fig_fmin_gam} the force sensitivity is plotted versus
the two-mode squeezing factor $s$, for four different values of
the damping rate, $\gamma = \{0.01, 0.1, 1, 10\}$ Hz, and again at
zero temperature. The dashed line is the SQL. As expected, the
force detection sensitivity worsens for increasing damping. As
shown by Fig.~\ref{Fig_fmin_t}, when $s=0$ and $\gamma=1$~Hz the
sensitivity is above the SQL. However it goes below the SQL in
correspondence of a two-mode squeezing coefficient $s\simeq1.5$ (a
squeezing of about $13$~dB). Interestingly enough, at fixed
$\gamma$, the minimum detectable force is not a monotonically
decreasing function of $s$, but it has a minimum, meaning that for
each $\gamma$ there is an optimal two-mode squeezing value
maximizing the force detection sensitivity. This feature was
lacking in Ref.~\cite{FMT04}, in which the best possible squeezing
was the highest one, and is a consequence of the inclusion of
damping and noise in the model. From a physical point of view,
this means that once that the interaction time $t_1$ is fixed, the
input entanglement and the dynamically generated entanglement
interfere in a nontrivial way, so that the optimal sensitivity is
achieved at a finite value of two-mode squeezing $s$.

\begin{figure}[h!]
\begin{center}
\includegraphics[width=0.45\textwidth]{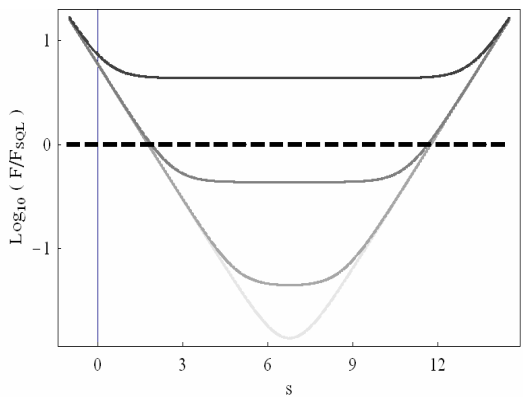}
\caption{Plot of
$\textrm{Log}_{10}(\textrm{F}/\textrm{F}_{\textrm{SQL}})$ versus the
two-mode squeezing factor $s$, at four different values of
temperature, $T = \{0, 0.03, 3, 300\}$~K, corresponding to
increasingly darker grey curves. The dashed line represents the SQL.
The other parameters are given by Table~1.} \label{Fig_fmin_tem}
\end{center}
\end{figure}
Finally we study the temperature dependence of the sensitivity of
the detection scheme. In Fig.~\ref{Fig_fmin_tem} we show the
behavior of the minimum detectable force versus $s$, at four
different values of the temperature, $T = \{0, 0.03, 3, 300\}$~K
and at fixed damping, $\gamma=1$ Hz, while the other parameters
are again those reported in Table~\ref{tab1}. We see that up to
mirror temperatures of the order of few Kelvin degrees, a nonzero
value of the input two-mode squeezing is able to compensate the
detrimental effects of damping and thermal noise acting on the
mirror and one can achieve a detection sensitivity better than the
SQL. This becomes impossible at room temperature, where the
minimum detectable force becomes larger than the SQL, for any
value of $s$.

\section{Conclusion}

We have studied in detail the optomechanical scheme for the
detection of weak forces proposed in Ref.~\cite{FMT04}, based on
the heterodyne measurement of a combination of two sideband modes
of an intense driving laser, scattered by a vibrational mode of a
highly reflecting mirror. In particular we have considered the
dynamical effects of damping and thermal noise acting on the
mirror vibrational mode, which were neglected in
Ref.~\cite{FMT04}. The dynamics of the system is characterized by
a bilinear coupling of the two optical sidebands with the
vibrational mode, which is able to generate significant
entanglement between the two sidebands for an appropriate duration
of the driving laser pulse~\cite{PM+03,FMT04}. This condition
corresponds to the highest signal-to-noise ratio for the detection
of a slowly varying mechanical force acting on the mirror which,
for extremely low values of the mirror damping and temperature,
can be better than the SQL for the detection of a force
\cite{BK92}. At more realistic values of damping and temperatures,
the minimum detectable force becomes larger than the SQL and the
dynamically generated entanglement is no more able to counteract
the effects of damping and thermal noise. However, we show that
the presence of additional entanglement in the input state of the
two sidebands may improve the performance of the detection scheme
and one can go significantly below the SQL, even in the presence
of non-negligible mirror damping and not too low temperatures. For
example the SQL can be beaten by adopting a vibrational mode with
a mechanical quality factor $\Omega/\gamma \simeq 10^7$ and at
temperatures around $3$ K.

This work has been supported by the European Commission through
the Integrated Project `Scalable Quantum Computing with Light and
Atoms' (SCALA), Contract No 015714, `Qubit Applications' (QAP),
Contract No 015848, funded by the IST directorate, and the
Ministero della Istruzione, dell'Universit\`{a} e della Ricerca
(PRIN-2005024254 and FIRB-RBAU01L5AZ).

\end{document}